\documentclass{ws-ijbc}
\usepackage{ws-rotating}     
\usepackage{graphicx}
\usepackage{epstopdf}
\begin{document}

\catchline{}{}{}{}{} 

\markboth{Karamanos et. al.}{Symbol-to-symbol correlation function at
the Feigenbaum point of the logistic map}

\title{Symbol-to-symbol correlation function at the Feigenbaum point of the logistic map\\}

\author{K. Karamanos}

\address{Complex Systems Group,Institute of Nuclear and Particle Physics, NRCPS Demokritos,GR 15310, Aghia Paraskevi
Attiki , Greece\\kkaraman@ulb.ac.be}

\author{I.S. Mistakidis}
\address{Hellenic Army Academy, Section of Physical Sciences and Applications,
Vari Attiki, Greece,\\is.mist@hotmail.com\\}

\author{S.I. Mistakidis}
\address{Section of Nuclear and Particle Physics, University of Athens,Panepistimiopolis, GR 15784,
Athens,Greece,\\smistakidis@phys.uoa.gr\\}

\maketitle

\date{\today}

\begin{abstract}
Recently, simple dynamical systems such as the 1-d maps on the interval,
gained significant attention in the context of statistical physics
and complex systems. The decay of correlations in these systems, can
be characterized and measured by correlation functions. In the
context of symbolic dynamics of the non-chaotic multifractal
attractors (i.e. Feigenbaum attractors), one observable, the
symbol-to-symbol correlation function, for the generating partition
of the logistic map, is rigorously introduced and checked with
numerical experiments. Thanks to the Metropolis-Stein-Stein (MSS)
algorithm this observable can be calculated analytically, giving
predictions in absolute accordance with numerical computations. The
deep, algorithmic structure of the observable is revealed clearly
reflecting the complexity of the multifractal attractor.

\end{abstract}

\section{Introduction}

Deterministic Chaos, discovered in  the beginning of the 20th
century, has been one of the basic subjects of Interdisciplinary
experimental, computational and theoretical research activity of the
end of the 20th century. Other similar topics include Fractals,
Finite automata, Cellular automata, Solitons, Linguistic dynamics,
Complex Systems, Ecosystems, Anticipatory Systems, General Systems
and Computer Assisted Music [Bai-Lin, 1994; Nicolis, 1991; Nicolis,
1995; Schr\"oder, 1991;
Schuster, 1984]. \\
The theory of Chaos is often highlighted as a theory of
unpredictability, dynamical randomness, irregular and erratic
behavior. Symbolic Dynamics has played an important role to Chaos
theory through Kolmogorov-Sinai entropy and Kolmogorov-Chaitin
Complexity.
A decisive step in symbolic dynamics has been done in [Metropolis et al., 1973]
where a new algorithm coming from the world of combinatorics, the
 Metropolis, Stein and Stein algorithm, has been used to generate and classify (through an abstract ordering)
the symbolic dynamics of the superstable orbits of unimodal maps. This led MSS to the Universal sequence
("U-sequence"), also referred to as "structural
 universality" in [Schuster, 1984].In a later step,one of the present authors (K.K.) and G. Nicolis have shown that the symbolic dynamics at the accumulation points of the $m \cdot {2^k}$ superstable orbits present self-similarity related to block-entropies. Precise Theorems  have been established and the relations to Fractals,
Turing Machines (Finite automata), and Transcendental Number Theory have been
 elucidated [Karamanos, 2001a].
A careful popular review to the subject of symbolic dynamics and
related topics has been given in a Lecture Note [Karamanos, 2000].\\
After important contributions by Feigenbaum [Feigenbaum, 1978;
Feigenbaum, 1979], and a wealth of related works the period-doubling
root to chaos is by now well-understood. Many authors have pointed
out similar trends between the transition to the non-chaotic
multifractal attractor (also called Feigenbaum attractor) and the
theory of second order phase transitions (Critical phenomena). A
related question in this context is about the analytic form of the
correlation function of the map. Motivated by the call of [Ruelle,
1986], J. Nicolis and Ebeling-Nicolis important publications paved
the road towards this direction [Nicolis \& Gaspard, 1994; Ebeling
 \& Nicolis, 1991; Ebeling \& Nicolis, 1992; Freund et al., 1996;
Nicolis et al., 1983; Rateitschak et al., 1996]. However, this task
had not yet been accomplished. After work by one of the present
authors (K.K.) and G. Nicolis [Karamanos \& Nicolis, 1999] it has
been shown that in the basis of the Metropolis, Stein and Stein
(MSS) algorithm, the symbolic dynamics at the Feigenbaum point is
universal,irrespectively of the detailed form of every unimodal map.
From this viewpoint it is interesting to study another type of
correlation function (observable), namely the symbol-to-symbol
correlation function for the logistic map. In this paper, we
formulate in a systematic basis and calculate
analytically all values of the symbol-to-symbol correlation function for the logistic map.\\
The self-similarity found by these authors, implies that the
invariant measure lives on a fractal set, which further connects
these results with work by Grassberger  and Mandelbrot [Grassberger,
1986; Karamanos \& Nicolis, 1999] in order to detect self-similarity
of symbolic dynamics, had used a modified block-entropy,
called  "block entropy by lumping".\\
Later on, [Karamanos, 2001b; Karamanos \&  Kotsireas, 2002] has
explored the limits of these methods, showing that block-entropy
analysis by lumping could be used for the detection of finite
automata in general. An entropy diagnostic for automaticity has been
announced in the international conference CASYS 2000 in Liege.
Generalizations of these results have circulated in a series of
papers. A completely mechanized
environment for large scale block entropy computations for arbitrary symbolic sequences has developed thanks to I. Kotsireas, in Maple. \\
Coming back now to the digital approach, from symbolic dynamics we
form some new constants and the algebraic nature of these constants
can be directly related to the Lyapunov exponent of the chaotic map.
In a first stage, the proof uses a relatively recent theorem by
[Allouche  \& Zamboni, 1998]. However a general proof that automatic
numbers are never algebraic irrational
 has seen the light of the day thanks to [Adamczewski \& Bugeaud, 2007]. These
theorems opened the road to the constructions
 of "super-universal" constants for any unimodal maps.\\
The study of correlation functions is the main subject of the
present paper. The paper is articulated as follows: In Sec. 2 we
present the part of the Metropolis, Stein and Stein (MSS) algorithm
which gives the universal symbolic dynamics at the Feigenbaum point
of unimodal maps. In Sec. 3 the self-similarity properties of this
symbolic dynamics under lumping are explored. In Sec. 4 we present
the connection with gliding. In Sec. 5  we introduce the notion of
the correlation functions. In Sec.  6 we accomplish our numerical
experimentation. In Sec. 7 we propose the extraction of the values
for the symbol-to-symbol correlation function at the accumulation
point of the ${2^k}$ superstable cycles. In Sec. 8 we draw the main
conclusions. In the next Sec we introduce symbolic dynamics with
particular emphasis to the symbolic dynamics at the Feigenbaum point
of unimodal maps.

\section{Coarse-graining and Universal symbolic dynamics of unimodal maps}

To produce symbolic dynamics out of the evolution of a given
dynamical system we set up a coarse-grained description
incorporating from the very beginning the idea that a physically
accessible state corresponds to a finite region rather than to a
single point of phase space. Let V be the set of cells in phase
space constituted by these regions assumed to be connected and
non-overlapping. As time goes on the phase space trajectory performs
transitions between cells thereby generating sequences of V symbols
which may be regarded as the letters of an alphabet. \\
For unimodal maps, the coarse-graining can be done in
a non-ambiguous ("canonical") manner, so that it is useful to
summarize some key properties of unimodal maps, defined as follows:

 i. They are functions ${f_\lambda }:[0,1] \to [0,1]$ depending on a real parameter ,$\lambda $,  (control parameter),
\\ii. ${f_\lambda }$ is continuous and at least piecewise differentiable in [0,1],
\\iii. ${f_\lambda }$ is convex and has a unique maximum but is otherwise arbitrary.\\
Let c be the position of the maximum, that is, if $x \ne c: f(x) <
f(c), f'(x) > 0$ if $x < c$ and $f'(x) < 0$  if  $x > c$. A particular
iterate, ${x_n}$  will be said to be of '0'or of L type and of '1'or
of R type,according to whether $\begin{array}{*{20}{c}}{{x_n} <
c}&{or}&{{x_n} > c}\end{array}$ , respectively. Given an initial
condition,${x_0}$ the 'minimum distinguishing information' about the
sequence of iterates ${x_n}$ will consist of a pattern of  0's (L's)
and 1's  (R's). Note that there are many possible ways to partition
the phase space and many candidates as initial points.\\ However,
according to the Julia theorem [Julia, 1918], the partition $[0,c]
\cup (c,1]$ and the initial point ${x_0} = c$  both turn out to be
the most fruitful, from the point of view of the extraction of
information available by the system. More specifically, the
resulting coarse-grained path (symbolic dynamics) is in a one-
to-one correspondence with the actual trajectory and, hence, there is
no loss of information in the topological sense. We are thus in
position to completely characterize the dynamical system as an
information generator by examining its symbolic dynamics only.This
is why this partition is called a generating one. For polynomial
maps, a stronger result called the 'Fatou and Julia Theorem' holds:
for appropriate $\lambda 's$ one can further localize the stable
periodic orbits starting from the critical point c (where $f'(c) = 0$) and iterating it.\\
Based on this theoretical framework, one can develop a whole
machinery for the construction of the 'patterns' of the
(super)stable periodic orbits of any period. This task has been
accomplished first in [Metropolis et al., 1973] for finite limit
sets and later completed in [Derrida et al., 1978], in order
 to include infinite limit sets. In view of the particular importance of the period-doubling route to chaos,
several authors tried to examine the symbolic dynamics of the corresponding attractors and this task has been achieved
in [Karamanos \& Nicolis, 1999] on a systematic basis.\\
Let P be the pattern (an abreviation of the symbolic trajectory)
associated with the superstable orbit of period-m (briefly called
here m-period). By definition,the (first)harmonic of P is the
pattern, $\widehat H(P) = P\mu P$ where $\mu$=L if P   contains an
odd number of R's and $\mu$=R otherwise. The procedure can be
iterated, so that one may speak of the second, third, $\ldots$, j-th
harmonic, etc., hereafter denoted as ${\widehat H^j}(P)$. MSS also
introduce the $\widehat H$-extension of a pattern P as the pattern P
generated by iterating the harmonic construction applied j times to
P,  when j increases indefinitely. In the
sequel, we shall rather adopt the notation ${\widehat {\rm H}^\infty }(P)$ for this asymptotic pattern.\\
MSS prove that, if P is allowed, $\widehat H(P)$ allowed too (their
Theorem 1). Furthermore,in their universal ordering, we have $P <
\widehat H(P)$ and the harmonics are adjacent, that is, no allowed
sequence exists between P and $\widehat H(P)$. Upon iterating the
process:

\begin{equation}
\label{eq:1}\forall j, P < \widehat H(P) < {\widehat H^2}(P) < ... <
{\widehat H^j}(P)
\end{equation}

For later use, we list hereafter the  first few harmonics associated
with the ${2^k}$ superstable cycles.

The  ${2^\infty }$ cycle, is given by:
\\
\\
${\begin{array}{l} P \equiv R(\begin{array}{*{20}{c}} 2&{period}
\end{array}) \to \widehat H(R) = \begin{array}{*{20}{c}}
{RLR}&{(4}&{period)}
\end{array} \to \\
{\widehat H^2}(R) = RLRRRLR = \begin{array}{*{20}{c}}
{RL{R^3}LR}&{(8}&{period) \to }
\end{array}\\
{\widehat H^3}(R) = RLRRRLRLRLRRRLR = \begin{array}{*{20}{c}}
{RL{R^3}LRLRL{R^3}LR}&{(16}&{period) \to }
\end{array}\\
{\widehat H^4}(R) = RLRRRLRLRLRRRLRRRLRRRLRLRLRRRLR = \\
\begin{array}{*{20}{c}}
{RL{R^3}LRLRL{R^3}L{R^3}L{R^3}LRLRL{R^3}LR}&{(32}&{period)}
\end{array}
\end{array}}$
\\

The logistic map is the archetype of a Complex System. Let us
elaborate. We introduce the logistic map in its familiar form :

\begin{equation}
\label{eq:2}{x_{n + 1}} = r{x_n}(1 - {x_n})
\end{equation}

For the logistic map in this form the generating partition is easily
computed, following the above-mentioned argument dating back to the
French Mathematician Gaston Julia. To be more specific,for $f(x) = r
\cdot x \cdot (1 - x)$ the equation $f'(c) = 0$ gives c=0.5 so that
the partition of the phase space (which in this case is the unit
interval I=[0,1] )  L=[0,0.5] and R=(0.5,1] is a generating
partition. As it has been already pointed out the information
content of the symbolic trajectory is the "minimum distinguishing
information" in the words of Metropolis et al. Needless to say, in
this representation the logistic map is viewed as an abstract
information generator.

\section{Self similarity and lumping}
For both reasons of completeness and for later use we compile here
some results first presented in [Karamanos \& Nicolis, 1999]. In
particular we establish certain invariance and self-similarity
properties of the symbolic sequences associated to the ${2^\infty }$
attractors. Taking advantage of these properties, we then introduce
a scheme mapping the original sequence into a new sequence of
hypersymbols with well-defined
statistical properties obtained by the lumping of groups of original symbols.\\
We define an operator, $\widehat K$, in the space of ${2^k}$
sequences (that is for the "patterns"  having the form ${\widehat
H^k}(R)$) through the following action which, in a sense, is inverse
to that of the harmonic operator  (we first fix a pattern $P =
{\widehat H^{m + 1}}(R)$):

i) Cut the last R of the pattern P (which necessarily ends by R and
has an odd number of symbols),

  ii) In the remaining part of the pattern, perform the lumpings :
  $\begin{array}{*{20}{c}}{RR \to L,}&{RL \to R}\end{array},$
  starting from the first R of the pattern from the left.
  Choosing $P = ({\widehat H^{m + 1}}(R))$ one can then show that

\begin{equation}
\label{eq:3}{\widehat K({\widehat H^{m + 1}}(R)) = {\widehat
H^m}(R)}
\end{equation}

Taking the limit, $m \to \infty $ in this relation, we have :
\begin{equation}
\label{eq:4}{\widehat K({\widehat H^\infty }(R)) = {\widehat
H^\infty }(R)}
\end{equation}
showing that the action of $\widehat {K}$ leaves the ${2^\infty }$
sequence invariant.The passage to this limit is essentially
equivalent to the existence of accumulation points and its validity
is established, for instance in [Feigenbaum, 1978; Feigenbaum, 1979]
and in another context, in [Daems  \&  Nicolis, 1994]. Actually this
self-similarity property can be extended straightforwardly to
guarantee invariance under a repetitive use of
 the operator $\widehat {K}$.
\begin{equation}
\label{eq:5}{{\widehat K^n}({\widehat H^\infty }(R)) = {\widehat
H^\infty }(R)}
\end{equation}
as long as n is finite. As an example, consider the subsequence
RLRRRLRL, which is part of the  symbolic trajectory. Applying the
rule twice, one obtains, successively, RLRR and RL, as
stipulated, precisely, in step (ii). Notice that this reduction would
be inapplicable,
 had the sequence been read by gliding one symbol each time.\\
In this way, we have explored the symmetries of the harmonic
operator. We shall now show that one can take profit of these
symmetries to generate the whole sequence. Indeed, we now express the
period-doubling symbolic sequence as a sequence generated by
substitutions instead of  being fixed point of the harmonic
operator. To this end, we introduce the morphism (substitution rule)
${{\mathop{\rm Re}\nolimits} ^{ - 1}}$  , defined as:
\begin{equation}
\label{eq:6}{\begin{array}{*{20}{c}} {{{{\mathop{\rm Re}\nolimits}
}^{ - 1}}(R) = RL},  &{{{{\mathop{\rm Re}\nolimits} }^{ - 1}}(L) =
RR}
\end{array}}
\end{equation}
And we observe that:
\begin{equation}
\label{eq:7}{\widehat H(R) = RLR = {{\mathop{\rm Re}\nolimits} ^{ -
1}}(R)R}
\end{equation}
By repetitive use of the above equation we find
\begin{equation}
\label{eq:8}{{\widehat H^n}(R) = [{({{\mathop{\rm Re}\nolimits} ^{ -
1}})^n}(R)][{({{\mathop{\rm Re}\nolimits} ^{ - 1}})^{n -
1}}(R)]...[{{\mathop{\rm Re}\nolimits} ^{ - 1}}(R)]R}
\end{equation}
Taking now the limit of this relation when $n \to \infty $ we find
asymptotically:
\begin{equation}
\label{eq:9}{{\widehat H^\infty }(R) = {({{\mathop{\rm Re}\nolimits}
^{ - 1}})^\infty }(R)}
\end{equation}
which is the desired relation, as it expresses the ${2^\infty }$
Feigenbaum sequence as the fixed point of the morphism
 ${{\mathop{\rm Re}\nolimits} ^{ - 1}}$ (eq. (6)) which is of constant length 2.
Operationally this guarantees that the
 successive replacements:

$\begin{array}{l}
R \to RL \to RLRR \to RLRRRLRL \to \\
 \to RLRRRLRLRLRRRLRR...
\end{array}$

generate indeed in each step the first part of the ${2^\infty }$
sequence. Due now to a theorem by A. Cobham, a sequence generated by
a set of substitution rules of length m, one for each letter
(morphism) can be in an equivalent manner generated by a
deterministic finite automaton (the lowest order Turing machine)
with m-states. (This is why the term "automatic").\\
We next proceed to reproduce the explicit evaluation of the
probability mass of the symbols  R and L. According to the MSS
algorithm,the number of letters of a permissible word increase from
construction steps $\nu$ to $\nu+1$ as ${\omega_{\nu  + 1}} = 2{\omega_\nu } +
1$ (with ${\omega_1} = 1$ ). The solution of this recurrence is ${\omega_\nu }
= {2^\nu } - 1$. On the other hand for the number of  R's, one has
($\omega_1^R = 1$)

\begin{equation}
\label{eq:10}{\omega_{\nu  + 1}^R = \left\{ {\begin{array}{*{20}{c}}
{\begin{array}{*{20}{c}} {2\omega_\nu ^R,}&{\omega_\nu ^R}&{odd}
\end{array}}\\
{\begin{array}{*{20}{c}} {2\omega_\nu ^R + 1,}&{\omega_\nu ^R}&{even}
\end{array}}
\end{array}} \right.}
\end{equation}

Whose solution is:

\begin{equation}
\label{eq:11}{\omega_\nu ^R = \left[ {\frac{{{2^{\nu  + 1}} - 1}}{3}}
\right],}
\end{equation}

It follows that the fraction of R's is:

\begin{equation}
\label{eq:12}{\begin{array}{*{20}{c}} p_{R = }&{\mathop
{\lim }\limits_{\nu  \to \infty } }
\end{array}\left( {\frac{{\left[ {\frac{{{2^{\nu  + 1}} - 1}}{3}} \right]}}{{{2^\nu } - 1}}} \right) = \frac{2}{3}}
\end{equation}

In other words, the {\it a priori}  probabilities of R and L in a long
sequence are:

\begin{equation}
\label{eq:13}{\begin{array}{*{20}{c}} {{p_R} = \frac{2}{3},}&{{p_L}
= \frac{1}{3}}&{(\begin{array}{*{20}{c}} {{2^\infty }}&{sequence}
\end{array})}
\end{array}}
\end{equation}

In the next Section we shall establish the connection with gliding.

\section{Further properties of frequencies for the ${2^\infty }$  sequence. Connection with Gliding.}

In [Karamanos \& Nicolis, 1999], it has been shown that one can
reduce, the study of the occurrence of words of an even length to
the study of words of an odd length, when reading starts from the
beginning of the sequence. The main idea for this computation is to
explicitly write down effective transition probabilities from first
principles. We shall
reproduce briefly the idea exposed in [Karamanos \& Nicolis, 1999] and give some examples.\\
The above analysis also provides information on the transition
probability matrix of the (coarse-grained) state vector. For
instance, in the case of the ${2^\infty }$ sequence, there is no $L
\to L$ transition, and the $R \to L$ or $R \to R$ transitions are
equally frequent. The 1-step transition matrix is thus

\begin{equation}
\label{eq:14}{W = \left( {\begin{array}{*{20}{c}}
{1/2}&{1/2}\\
1&0
\end{array}} \right)}
\end{equation}

Suppose we begin with a symbol R, randomly chosen on the ${2^\infty}$ sequence.
This has a probability 2/3 to happen. The next symbol
has 1/2 probability  to be an R and equal probability to an L, in
view of the 1-step transition matrix eq.(14). For the next step, we
have the following, keeping in mind the replacement rules of eq.(6):

i)  the configuration, RR has a probability 1/2 to continue as RRR and a probability 1/2 to continue as RRL.\\
ii) the configuration, RL, continues certainly (with probability 1)
with R, to form RLR, as we already know from eq. (14).

If the subsequence begins now with an L, the situation is slightly
more complicated. We first observe that the operator $\widehat {\rm
K}$ introduced as a kind of inverse of the harmonic operator
previously , is not uniquely defined. One could, for instance,
define an operator $\widehat {K'}$ which cuts the first R of the
pattern, $P = {\widehat H^q}$ and performs the replacements ,
$\begin{array}{*{20}{c}}{RR \to L,}&{LR \to R}\end{array}$ to the
rest of the sequence. Then it is easy to show that  $\widehat {K'}$
is also a (partial) inverse of $\widehat H$ entailing that, in the
case of subsequences which start with an L, self-similarity is
manifested with (new) hypersymbols formulated under the replacements
$\begin{array}{*{20}{c}}{RR \to L,}&{LR \to R} \end{array}$.If we
begin so with an L, this has a probability 1/3 to happen. For the
next symbol, we are certain that it will be R (from eq. (14)), to
form LR. The latter configuration has a probability of 1/2 to
continue as LRR and an equal probability, 1/2, to continue as LRL.
Locally, one can also use the  conclusions of the previous section
as, for instance, the recurrence time distribution : a configuration $\ldots$LRR will continue, with certainty, with an R.\\
To conclude, we thus have five words of length three: three words
starting with an R, namely RRR, RRL and RLR, with corresponding
frequencies of occurrence, 1/6, 1/6, and 1/3, respectively, and two
words starting with an L, namely LRR and LRL, with corresponding
frequencies of occurrence, 1/6 and 1/6, respectively.\\
Excluding configurations which do not obey the basic requirement of
self-similarity, we proceed like this {\it ad infinitum}. Fig. 2  also
illustrates this point.

\begin{figure}[h]
\vspace{-15pt}
        \centering
                \includegraphics[width=0.60\textwidth]{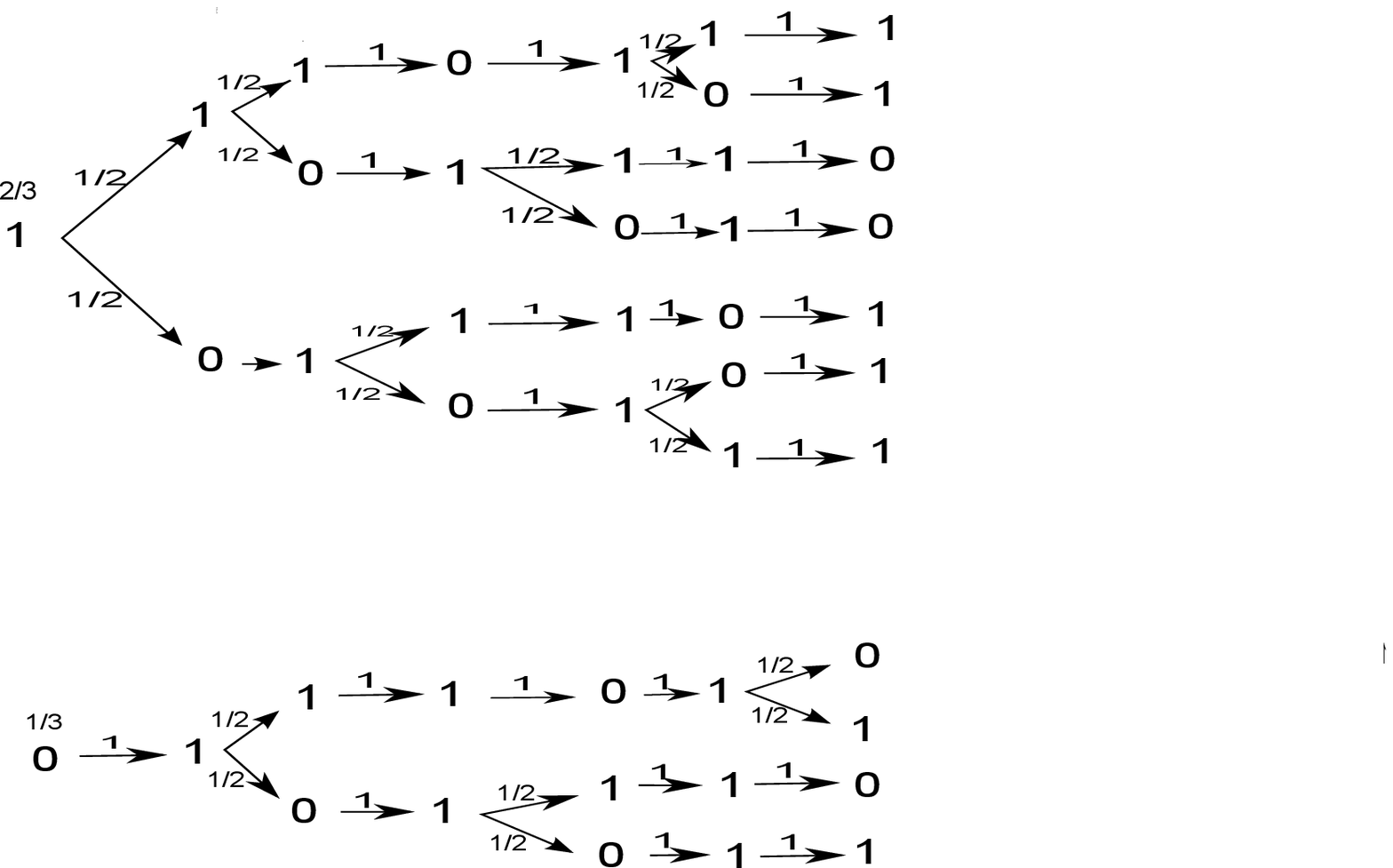}
                \vspace{-0.3cm}
                \caption{This figure illustrates the mechanism of generation of the words encountered when reading the sequence by
                gliding.}
\end{figure}

This procedure allows to write down all the words of an odd length
encountered when gliding on the ${2^\infty }$ sequence, as well as
their frequencies. Using the effective n-step (gliding) transition
probabilities just introduced, we are in the position to write down
explicitly the entropies for words of an odd length in a
constructive manner (that is, step-by-step). In the next section we
introduce the main concern  of this paper, different types of
correlation functions.

\section{Correlation functions}

First of all we define the correlation function of the trajectory as:

\begin{equation}
\label{eq:15}{C_{un}(m) = \mathop {\lim }\limits_{N \to \infty }
\frac{1}{N}\sum\limits_{i = 0}^{N - 1} {{{\widehat x}_{i +
m}}{{\widehat x}_i}}, where \begin{array}{*{20}{c}} {{{\widehat
x}_i} = {f^i}({x_0}) - \overline x },&{\overline x }
\end{array} = \mathop {\lim }\limits_{N \to \infty } \frac{1}{N}\sum\limits_{i = 0}^{N - 1} {{f^i}({x_0})}}
\end{equation}

In direct analogy with the un-normalized correlation function we
also introduce here the normalized correlation function:
\begin{equation}
\label{eq:16}C(m) = \frac{{{C_{un}}(m)}}{{{C_{un}}(0)}} =
\frac{{{C_{un}}(m)}}{{{\sigma ^2}}}
\end{equation}
where $\sigma $ is the mean standard deviation.

 From these definitions follows that $C_{un}(m)$ yields another measure for the
irregularity of the sequence of iterates ${x_0}$ ,$f({x_0})$,
${f^2}({x_0})$ $\ldots$.It tells us how much the deviations of the
iterates from their average value, ${\widehat x_i} = {x_i} -
\overline x$ that are m steps apart (i. e., ${\widehat x_{i + m}}$
and ${\widehat x_i}$) "know" about each other, on the average. If
$C(m)\nrightarrow 0$ as $m \to \infty $ then the system does not
have the mixing property. The problem of determining the correlation
function of an arbitrary dynamical system is intractable in the
general case. This is the reason to resort to other computable
observables such as the symbol-to-symbol correlation function [Daems
\& Nicolis, 1994]. \\
So,in analogy with the correlation function of the trajectory we can
introduce the symbol-to-symbol correlation function as:

\begin{equation}
\label{eq:17}{K_{un}(m) = \mathop {\lim }\limits_{N \to \infty }
\frac{1}{N}\sum\limits_{i = 0}^{N - 1} {{{\widehat y}_{i +
m}}{{\widehat y}_i}}}
\end{equation}

where ${\widehat y_i}$=0,1 when ${x_i} \leqslant 0.5$ or ${x_i} >
0.5$ respectively and

\begin{equation}
\label{eq:18}{\begin{array}{*{20}{c}}
  {{{\widehat y}_i} = y({f^i}({x_0})) - \overline y }&{where}&{\overline y  = }
\end{array}\mathop {\lim }\limits_{N \to \infty } \frac{1}{N}\sum\limits_{i = 0}^{N - 1} {y({f^i}({x_0}))}}
\end{equation}

In the same way we can define the normalized symbol-to-symbol
correlation function

\begin{equation}
\label{eq:19}{\rm K}(m) = \frac{{{{\rm K}_{un}}(m)}}{{{{\rm
K}_{un}}(0)}} = \frac{{{{\rm K}_{un}}(m)}}{{{{\sigma '}^2}}}
\end{equation}

\section{Numerical experimentation}

Motivated by previous work on correlation functions [Daems  \&
Nicolis, 1994; Ruelle, 1986], we explore here numerically the
properties of the symbol-to-symbol
correlation function.\\
In order to cope with the problem of the analytic form of
correlation functions we stalled certain number of numerical
experiments. For the logistic map at the Feigenbaum point
r=3.56994567$\ldots$ we have calculated the un-normalized
symbol-to-symbol correlation function for $50 \cdot {10^6}$
iterations (we eliminated the first ${10^5}$ iterations to avoid
transients), starting from the initial point ${x_0} = 0.5$ As also
stated in [Karamanos \& Nicolis, 1999],
exactly at the Feigenbaum point the Lyapunov exponent strictly vanishes $\lambda$= 0.\\
We now turn back to the structure of the un-normalized
symbol-to-symbol correlation function as it is found numerically for
$50 \cdot {10^6}$  iterations (this scheme is depicted in Fig.2):

\begin{equation}
\label{eq:20}{K_{un}}(m) = \left\{ {\begin{array}{*{20}{c}}
{ - \frac{1}{9},}&{m - odd = 1 + 2 \cdot k}&{(k = 0,1,2,...)}\\
{\frac{1}{{18}},}&{m = 2 + 4 \cdot k}&{(k = 0,1,2,...)}\\
{\frac{1}{{7.2}},}&{m = 4 + 8 \cdot k}&{(k = 0,1,2,...)}\\
{\frac{1}{{5.537}},}&{m = 8 + 16 \cdot k}&{(k = 0,1,2,...)}\\
{\sim\frac{1}{5},}&{m = 16 + 32 \cdot k}&{(k = 0,1,2,...)}\\
{\sim\frac{1}{{4.720}},}&{m = 32 + 64 \cdot k}&{(k = 0,1,2,...)}\\
{\sim\frac{1}{{4.6085}},}&{m = 64 + 128 \cdot k}&{(k = 0,1,2,...)}\\
{\sim\frac{1}{{4.5521}},}&{m = 128 + 256 \cdot k}&{(k = 0,1,2,...)}\\
{\sim\frac{1}{{4.5271}},}&{m = 256 + 512 \cdot k}&{(k = 0,1,2,...)}\\
{\sim\frac{1}{{4.5121}},}&{m = 512 + 1024 \cdot k}&{(k = 0,1,2,...)}\\
{\sim\frac{1}{{4.5072}},}&{m = 1024 + 2048 \cdot k}&{(k =
0,1,2,...)}
\end{array}} \right.
\end{equation}

In a more compact form we can announce the following empirical rule
for the symbol-to-symbol correlation function (observable):

\begin{equation}
\label{eq:21}{K_{un}(m) = {A_r} \cdot {\delta _{m,{2^{r - 1}} \cdot
(1 + 2 \cdot k)}}}
\end{equation}
where for a given (fixed)  r, $r \in \{ 1,2,3,....\}$ , ${A_r}$   is
a constant depending only on r , and k takes all the values from
the set of natural numbers $\{ 0,1,2,3,....\}$ . As we will show in Section 7 one can determine the constants ${A_r}$ from
first principles.\\
As a final comment we observe that according to our computations
that for $m \ge 256$ the correlation function ($K_{un}(m)$)
approaches the value 2/9.

\begin{figure}[h]
\vspace{-15pt}
        \centering
                \includegraphics[width=0.60\textwidth]{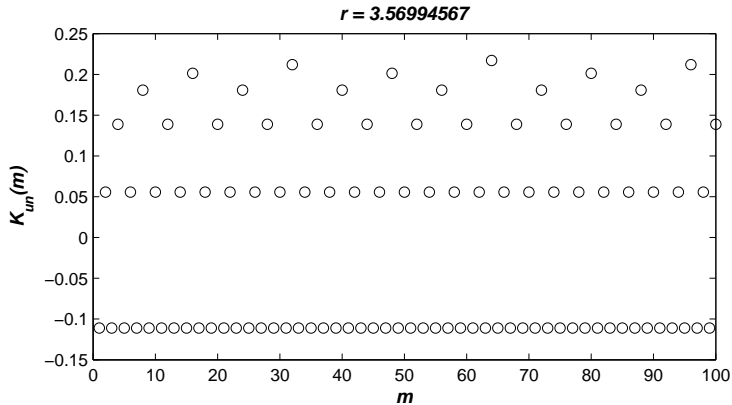}
                \vspace{-0.3cm}
                \caption{The symbol-to-symbol correlation function is depicted for the logistic map exactly at the accumulation point r = FP and
                with initial condition ${x_0} = 0.5$. The first  ${10^5}$  iterations have been eliminated in order to
                exclude transients, and the subsequent  $50 \cdot {10^6}$  iterations have been taken into
                account for the calculations. The open circles in the diagram represent the values of the
                symbol-to-symbol correlation function as a function of the separation steps m. The experimentally determined Lyapunov
                 is $ - 5,94 \cdot {10^{ - 5}}.$}
\end{figure}

\section{Analytic calculation of the symbol-to-symbol correlation function}

From the Metropolis, Stein and Stein algorithm we find that the
first terms of the ${2^\infty }$ symbolic sequence are:
\\ RLRRRLRLRLRRRLRRRLRRRLRLRLRRRLR$\ldots$ denoted by $\mathbb{A}$\\
Corresponding to this symbolic sequence is the arithmetic sequence :
 \\ 1011101010111011101110101011101$\ldots$ denoted by ($\mathbb{S}$)\\

 Starting from the initial point ${x_0} = 0.5$.
Our interest is to understand the correlations of the above
arithmetic sequences.One way is to examine the corresponding
symbol-to-symbol correlation function.\\
It is useful to introduce now some definitions and compact auxiliary
notation  in connection with arithmetic sequences. Let ${\left(
{{a_n}} \right)_{n \in N}} = {a_1},{a_2},...,{a_n}$ and ${\left(
{{b_n}} \right)_{n \in N}} = {b_1},{b_2},...,{b_n}$ be two
arithmetic sequences.We define the following composition law:

\begin{equation}
\label{eq:22}{{a_N} \circ {b_N} = \frac{1}{N}\left( {{a_1} \cdot
{b_1} + {a_2} \cdot {b_2} + ... + {a_N} \cdot {b_N}} \right)}
\end{equation}
and taking the limit (if it exists) as $N \to \infty$ we define $\mathbb{A} \circ \mathbb{{\ss}} = \mathop {\lim }\limits_{N \to \infty } \frac{1}{N}\sum\limits_{i = 1}^N {\left( {{a_i} \cdot {b_i}} \right)}$ .\\
 It is easy to show that the
mean value of  ($\mathbb{S}$) is

\begin{equation}
\label{eq:23}{ < y >  = \frac{2}{3}R + \frac{1}{3}L = \frac{2}{3}1 +
\frac{1}{3}0 = \frac{2}{3}}
\end{equation}

So that
\begin{equation}
\label{eq:24}{{\left( { < y > } \right)^2} = \frac{4}{9}}
\end{equation}

The subsequence of ($\mathbb{S}$ ) corresponding to the even indices
($2\rho $) is :
\\010001010100010$\ldots$ denoted by $\overline{\mathbb{S}}$
\\The subsequence of ($\mathbb{S}$ ) corresponding to the odd indices ($2\rho  + 1$) is:
\\1111111111111111$\ldots$ denoted by $\mathbb{I}$
\\In the Appendix A we shall prove that sequence ($\overline{\mathbb{S}}$ ) is the complementary of the sequence ( $\mathbb{S}$) , (that it is generated from the
sequence ( $\mathbb{S}$) by the replacement
$\begin{array}{*{20}{c}}{0 \to 1}&,&{1 \to 0}\end{array}$) .We shall
also prove that sequence ($\mathbb{I}$) is periodic of period 1. \\
We now introduce the following notation:

\begin{equation}
\label{eq:25}{{\mathbb{A}_q} = {a_q} \cdot {a_{q + 1}} \cdot ....
\cdot {a_{q + n}}}
\end{equation}

In the sequel we shall show the following proposition:

\text{\bf LEMMA 1}:  For the odd indices of the correlation function
we have ${K_{un}}(m) =  - \frac{1}{9}$ , for $m = 1 + 2\rho$  (odd).
We first illustrate the above lemma for K(1) and K(3). We thereby
turn to the problem of determining K(1):

\begin{equation}
 \label{eq:26}{\begin{gathered}
  K(1) +  < y{ > ^2} = \mathop {\lim }\limits_{N \to \infty } \left[ {\frac{1}{{2N}}({{\widehat y}_1} \cdot {{\widehat y}_2} + {{\widehat y}_3} \cdot {{\widehat y}_4} + {{\widehat y}_5} \cdot {{\widehat y}_6} + ...)} \right] =
  \\= \mathop {\lim }\limits_{N \to \infty } \left[ {\frac{1}{2}\frac{1}{N}\sum\limits_{n = 0}^N {\left( {{{\widehat y}_{2n + 1}} \cdot {{\widehat y}_{2n + 2}} + {{\widehat y}_{2n + 3}} \cdot {{\widehat y}_{2n + 4}}} \right)} } \right] =  \hfill \\
   = \frac{1}{2}({\mathbb{S}_{2n + 1}} \circ {\mathbb{S}_{2n + 2}} + {\mathbb{S}_{2n + 3}} \circ {\mathbb{S}_{2n + 4}}) = \frac{1}{2}(\mathbb{I} \circ \overline{\mathbb{S}}  + \overline{\mathbb{S}}  \circ \mathbb{I}) = \frac{1}{2}\left( {1 \cdot \frac{1}{3} + \frac{1}{3}\cdot1} \right) = \frac{1}{3} \hfill \\
\end{gathered}}
 \end{equation}

So that,

\begin{equation}
 \label{eq:27}{K(1) = \frac{1}{3} -  < y{ > ^2} = \frac{1}{3} - \frac{4}{9} =  - \frac{1}{9}}
 \end{equation}

Next we compute K(3). We found that:

\begin{equation}
 \label{eq:28}{\begin{gathered}
  K(3) +  < y{ > ^2} = \mathop {\lim }\limits_{N \to \infty } \left[ {\frac{1}{{2{\rm N}}}({{\widehat y}_1} \cdot {{\widehat y}_4} + {{\widehat y}_2} \cdot {{\widehat y}_5} + {{\widehat y}_3} \cdot {{\widehat y}_6} + ...)} \right] =\\
   \mathop {\lim }\limits_{N \to \infty } \left[ {\frac{1}{2}\frac{1}{{\rm N}}\sum\limits_{\rho  = 0}^N {\left( {{{\widehat y}_{2\rho  + 1}} \cdot {{\widehat y}_{2\rho  + 4}} + {{\widehat y}_{2\rho  + 2}} \cdot {{\widehat y}_{2\rho  + 5}}} \right)} } \right] =  \hfill \\
   = \frac{1}{2}({\mathbb{S}_{2\rho  + 1}} \circ {\mathbb{S}_{2\rho  + 4}} + {\mathbb{S}_{2\rho  + 2}} \circ {\mathbb{S}_{2\rho  + 5}}) = \frac{1}{2}(\mathbb{I} \circ \overline{\mathbb{S}}  + \overline{\mathbb{S}}  \circ \mathbb{I}) = \frac{1}{2}\left( {1 \cdot \frac{1}{3} + \frac{1}{3}\cdot 1} \right) = \frac{1}{3} \hfill \\
\end{gathered}}
 \end{equation}

So:

\begin{equation}
 \label{eq:29}{K(3) = \frac{1}{3} -  < y{ > ^2} = \frac{1}{3} - \frac{4}{9} =  - \frac{1}{9}}
 \end{equation}

Thus we observe  ${\rm K}(1) = K(3) =  - \frac{1}{9}$.\\
We outline now the general proof of this lemma:

\begin{equation}
 \label{eq:30}{\begin{gathered}
  K(2\rho  + 1) +  < y{ > ^2} = \frac{1}{2}({\mathbb{S}_{2\nu  + 1}} \circ {\mathbb{S}_{2\nu  + 1 + 2\rho  + 1}} + {\mathbb{S}_{2\nu  + 2}} \circ {\mathbb{S}_{2\nu  + 2 + 2\rho  + 1}}) = \frac{1}{2}({\mathbb{S}_{2\nu  + 1}} \circ {\mathbb{S}_{2(\nu  + \rho ) + 2}} + {\mathbb{S}_{2\nu  + 2}} \circ {\mathbb{S}_{2(\nu  + \rho ) + 3}}) =  \hfill \\
   = \frac{1}{2}(\mathbb{I} \circ {\mathbb{S}_{2\lambda  + 2}} + \overline{\mathbb{S}}  \circ {\mathbb{S}_{2\rho  + 3}}) = \frac{1}{2}(\mathbb{I} \circ \overline{\mathbb{S}}  + \overline{\mathbb{S}}  \circ \mathbb{I}) = \frac{1}{3} \hfill \\
\end{gathered}}
 \end{equation}
so that:
\begin{equation}
 \label{eq:31}{K(2\rho+1) = \frac{1}{3} -  < y{ > ^2} = \frac{1}{3} - \frac{4}{9} =  - \frac{1}{9}}
 \end{equation}

\text{\bf LEMMA 2}: For the even indices of the form 2+4k we have
:${K_{un}}(2 + 4k) = \frac{1}{{18}}$ . Again we first illustrate the
above lemma for K(2) and K(4).

\begin{equation}
 \label{eq:32}{\begin{gathered}
  K(2) +  < y{ > ^2} = \mathop {\lim }\limits_{N \to \infty } \left[ {\frac{1}{{2{\rm N}}}({{\widehat y}_1} \cdot {{\widehat y}_3} + {{\widehat y}_2} \cdot {{\widehat y}_4} + ...)} \right] = \mathop {\lim }\limits_{N \to \infty } \left[ {\frac{1}{2}\frac{1}{{\rm N}}\sum\limits_{\rho  = 0}^N {\left( {{{\widehat y}_{2\rho  + 1}} \cdot {{\widehat y}_{2\rho  + 3}} + {{\widehat y}_{2\rho  + 2}} \cdot {{\widehat y}_{2\rho  + 4}}} \right)} } \right] =  \hfill \\
  \frac{1}{2}({\mathbb{S}_{2\nu  + 1}} \circ {\mathbb{S}_{2\nu  + 3}} + {\mathbb{S}_{2\nu  + 2}} \circ {\mathbb{S}_{2\nu  + 4}}) = \frac{1}{2}(\mathbb{I} \circ \mathbb{I} + 0) = \frac{1}{2} \cdot 1 = \frac{1}{2} \Rightarrow  \hfill \\
   \Rightarrow K(2) = \frac{1}{2} - \frac{4}{9} = \frac{1}{{18}} \hfill \\
\end{gathered}}
 \end{equation}

\begin{equation}
 \label{eq:33}{\begin{gathered}
  K(4) +  < y{ > ^2} = \mathop {\lim }\limits_{N \to \infty } \left[ {\frac{1}{{2{\rm N}}}({{\widehat y}_1} \cdot {{\widehat y}_3} + {{\widehat y}_2} \cdot {{\widehat y}_4} + ...)} \right] = \mathop {\lim }\limits_{N \to \infty } \left[ {\frac{1}{2}\frac{1}{{\rm N}}\sum\limits_{\rho  = 0}^N {\left( {{{\widehat y}_{2\rho  + 1}} \cdot {{\widehat y}_{2\rho  + 5}} + {{\widehat y}_{2\rho  + 2}} \cdot {{\widehat y}_{2\rho  + 6}}} \right)} } \right] =  \hfill \\
  \frac{1}{2}({\mathbb{S}_{2\rho  + 1}} \circ {\mathbb{S}_{2\rho  + 5}} + {\mathbb{S}_{2\rho  + 2}} \circ {\mathbb{S}_{2\rho  + 6}}) = \frac{1}{2}(\mathbb{I} \circ \mathbb{I} + \overline{\mathbb{S}}  \circ {\mathbb{S}_{2\rho  + 6}}) \simeq \frac{1}{2} \cdot (1 + \frac{1}{{10}}) = \frac{7}{{12}} \Rightarrow  \hfill \\
   \Rightarrow K(4) = \frac{7}{{12}} -  < y{ > ^2} = \frac{{10}}{{72}} \hfill \\
\end{gathered}}
 \end{equation}

Coming back to Fig.1 we are thus ready to achieve a spectacular
reduction of complexity: for the symbol-to-symbol correlation
function K(m), the study reduces to the calculation of the
1$\longrightarrow$$\ldots1$ probabilities, that is of the R$\ldots$R
contributions in the sum, as the R$\ldots$L , L$\ldots$L ,
L$\ldots$R  contributions vanish. In order to compute $K(m) +  < y{
> ^2}$ we multiply the correspondingly transition
probabilities in chain for the blocks R$\ldots$R .\\
We first recover the few initial values of $K(m) +  < y{ > ^2}$
(m=1,2,3,4) calculated previously from first principles starting
from the diagram of Fig.1 :

\begin{equation}
 \label{eq:34}{ \ \bullet K(1) +  < y{ > ^2} = \frac{2}{3}}
\end{equation}

 \begin{equation}
 \label{eq:35}{ \bullet K(2) +  < y{ > ^2} = \frac{2}{3} \cdot \frac{1}{2} \cdot \frac{1}{2} + \frac{2}{3} \cdot \frac{1}{2} \cdot 1 = \frac{1}{2}}
 \end{equation}

 \begin{equation}
 \label{eq:36}{ \bullet K(3) +  < y{ > ^2} = \frac{2}{3} \cdot \frac{1}{2} \cdot \frac{1}{2} + \frac{2}{3} \cdot \frac{1}{2} \cdot 1 \cdot \frac{1}{2} = \frac{1}{3}}
 \end{equation}

\begin{equation}
 \label{eq:37}{\begin{array}{l}
 \bullet K(4) +  < y{ > ^2} = \frac{2}{3} \cdot \frac{1}{2} \cdot \frac{1}{2} \cdot 1 \cdot 1 + \frac{2}{3} \cdot \frac{1}{2} \cdot \frac{1}{2} \cdot 1 \cdot \frac{1}{2} + \frac{2}{3} \cdot \frac{1}{2} \cdot 1 \cdot \frac{1}{2} \cdot 1 + \frac{2}{3} \cdot \frac{1}{2} \cdot 1 \cdot \frac{1}{2} \cdot 1 \Rightarrow \\
 \Rightarrow K(4) +  < y{ > ^2} = \frac{7}{{12}}
\end{array}}
 \end{equation}

We then proceed for the K(m) for the next few values of m:

\begin{equation}
\label{eq:38}{\begin{array}{l}
\bullet K(5) +  < y{ > ^2} = \frac{2}{3} \cdot \frac{1}{2} \cdot \frac{1}{2} \cdot 1 \cdot 1 \cdot \frac{1}{2} + \frac{2}{3} \cdot \frac{1}{2} \cdot \frac{1}{2} \cdot 1 \cdot \frac{1}{2} \cdot 1 + \frac{2}{3} \cdot \frac{1}{2} \cdot \frac{1}{2} \cdot 1 \cdot \frac{1}{2} \cdot 1 + \frac{2}{3} \cdot \frac{1}{2} \cdot 1 \cdot \frac{1}{2} \cdot 1 \cdot \frac{1}{2} \Rightarrow \\
\Rightarrow K(5) +  < y{ > ^2} = \frac{2}{3} \cdot \frac{1}{2} \cdot
\frac{1}{2} \cdot \frac{1}{2} \cdot \left( {1 + 1 + 1 + 1} \right) =
\frac{1}{3}
\end{array}}
\end{equation}

\begin{equation}
\label{eq:39}{\begin{array}{l}
\bullet K(6) +  < y{ > ^2} = \frac{2}{3}\left( {\frac{1}{2} \cdot \frac{1}{2} \cdot \frac{1}{2} + \frac{1}{2} \cdot \frac{1}{2} \cdot \frac{1}{2} + \frac{1}{2} \cdot \frac{1}{2} + \frac{1}{2} \cdot \frac{1}{2} \cdot \frac{1}{2} + \frac{1}{2} \cdot \frac{1}{2} \cdot \frac{1}{2}} \right) \Rightarrow \\
\Rightarrow K(6) +  < y{ > ^2} = \frac{2}{3}\left( {\frac{1}{2} +
\frac{1}{4}} \right) = \frac{1}{2}
\end{array}}
\end{equation}

 \begin{equation}
 \label{eq:40}{ \begin{array}{l}
 \bullet K(7) +  < y{ > ^2} = \frac{2}{3}(\frac{1}{2} \cdot \frac{1}{2} \cdot \frac{1}{2} + \frac{1}{2} \cdot \frac{1}{2} \cdot \frac{1}{2} + \frac{1}{2} \cdot \frac{1}{2} \cdot \frac{1}{2} + \frac{1}{2} \cdot \frac{1}{2} \cdot \frac{1}{2}) = \frac{2}{3} \cdot 4 \cdot \frac{1}{2} \cdot \frac{1}{2} \cdot \frac{1}{2} = \frac{1}{3} \Rightarrow \\
 \Rightarrow K(7) +  < y{ > ^2} = \frac{1}{3}
\end{array}}
 \end{equation}

 \begin{equation}
 \label{eq:41}{\begin{array}{l}
 \bullet K(8) +  < y{ > ^2} = \frac{2}{3}\left( {\frac{1}{2} \cdot \frac{1}{2} \cdot \frac{1}{2} + \frac{1}{2} \cdot \frac{1}{2} \cdot \frac{1}{2} + \frac{1}{2} \cdot \frac{1}{2} \cdot \frac{1}{2} \cdot \frac{1}{2} + \frac{1}{2} \cdot \frac{1}{2} \cdot \frac{1}{2} + \frac{1}{2} \cdot \frac{1}{2} \cdot \frac{1}{2} + \frac{1}{2} \cdot \frac{1}{2} \cdot \frac{1}{2} + \frac{1}{2} \cdot \frac{1}{2} \cdot \frac{1}{2}} \right) \Rightarrow \\
 \Rightarrow K(8) +  < y{ > ^2} = \frac{{15}}{{24}}
\end{array}}
 \end{equation}

\begin{equation}
 \label{eq:42}{\begin{array}{l}
 \bullet K(9) +  < y{ > ^2} = \frac{2}{3}\left( {\frac{1}{2} \cdot \frac{1}{2} \cdot \frac{1}{2} \cdot \frac{1}{2} + \frac{1}{2} \cdot \frac{1}{2} \cdot \frac{1}{2} + \frac{1}{2} \cdot \frac{1}{2} \cdot \frac{1}{2} \cdot \frac{1}{2} + \frac{1}{2} \cdot \frac{1}{2} \cdot \frac{1}{2} \cdot \frac{1}{2} + \frac{1}{2} \cdot \frac{1}{2} \cdot \frac{1}{2} + \frac{1}{2} \cdot \frac{1}{2} \cdot \frac{1}{2} \cdot \frac{1}{2}} \right) \Rightarrow \\
 \Rightarrow K(9) +  < y{ > ^2} = \frac{1}{3}
\end{array}}
 \end{equation}

 \begin{equation}
 \label{eq:43}{\begin{array}{l}
 \bullet K(10) +  < y{ > ^2} =\\ \frac{2}{3}\left( {\frac{1}{2} \cdot \frac{1}{2} \cdot \frac{1}{2} \cdot \frac{1}{2} + \frac{1}{2} \cdot \frac{1}{2} \cdot \frac{1}{2} \cdot \frac{1}{2} + \frac{1}{2} \cdot \frac{1}{2} \cdot \frac{1}{2} + \frac{1}{2} \cdot \frac{1}{2} \cdot \frac{1}{2} + \frac{1}{2} \cdot \frac{1}{2} \cdot \frac{1}{2} + \frac{1}{2} \cdot \frac{1}{2} \cdot \frac{1}{2} + \frac{1}{2} \cdot \frac{1}{2} \cdot \frac{1}{2} \cdot \frac{1}{2} + \frac{1}{2} \cdot \frac{1}{2} \cdot \frac{1}{2} \cdot \frac{1}{2}} \right) \Rightarrow \\
 \Rightarrow K(10) +  < y{ > ^2} = \frac{1}{2}
\end{array}}
 \end{equation}

 \begin{equation}
 \label{eq:44}{\begin{array}{l}
 \bullet K(11) +  < y{ > ^2} = \frac{2}{3}\left( {\frac{1}{2} \cdot \frac{1}{2} \cdot \frac{1}{2} \cdot \frac{1}{2} + \frac{1}{2} \cdot \frac{1}{2} \cdot \frac{1}{2} \cdot \frac{1}{2} + \frac{1}{2} \cdot \frac{1}{2} \cdot \frac{1}{2} + \frac{1}{2} \cdot \frac{1}{2} \cdot \frac{1}{2} \cdot \frac{1}{2} + \frac{1}{2} \cdot \frac{1}{2} \cdot \frac{1}{2} + \frac{1}{2} \cdot \frac{1}{2} \cdot \frac{1}{2} \cdot \frac{1}{2}} \right) \Rightarrow \\
 \Rightarrow K(11) +  < y{ > ^2} = \frac{1}{3}
\end{array}}
 \end{equation}

 \begin{equation}
 \label{eq:45}{\begin{array}{l}
 \bullet K(12) +  < y{ > ^2} = \frac{2}{3}(\frac{1}{2} \cdot \frac{1}{2} \cdot \frac{1}{2} \cdot \frac{1}{2} + \frac{1}{2} \cdot \frac{1}{2} \cdot \frac{1}{2} \cdot \frac{1}{2} + \frac{1}{2} \cdot \frac{1}{2} \cdot \frac{1}{2} +\\ \frac{1}{2} \cdot \frac{1}{2} \cdot \frac{1}{2} \cdot \frac{1}{2} + \frac{1}{2} \cdot \frac{1}{2} \cdot \frac{1}{2} \cdot \frac{1}{2} + \frac{1}{2} \cdot \frac{1}{2} \cdot \frac{1}{2} \cdot \frac{1}{2} + \frac{1}{2} \cdot \frac{1}{2} \cdot \frac{1}{2} \cdot \frac{1}{2}\\
 + \frac{1}{2} \cdot \frac{1}{2} \cdot \frac{1}{2} + \frac{1}{2} \cdot \frac{1}{2} \cdot \frac{1}{2} \cdot \frac{1}{2} + \frac{1}{2} \cdot \frac{1}{2} \cdot \frac{1}{2} \cdot \frac{1}{2} + \frac{1}{2} \cdot \frac{1}{2} \cdot \frac{1}{2} \cdot \frac{1}{2} + \frac{1}{2} \cdot \frac{1}{2} \cdot \frac{1}{2} \cdot \frac{1}{2}) \Rightarrow K(12) +  < y{ > ^2} = \frac{7}{{12}}
\end{array}}
 \end{equation}

\begin{equation}
 \label{eq:46}{\begin{array}{l}
 \bullet K(13) +  < y{ > ^2} = \frac{2}{3}(\frac{1}{2} \cdot \frac{1}{2} \cdot \frac{1}{2} \cdot \frac{1}{2} + \frac{1}{2} \cdot \frac{1}{2} \cdot \frac{1}{2} \cdot \frac{1}{2} + \frac{1}{2} \cdot \frac{1}{2} \cdot \frac{1}{2} \cdot \frac{1}{2} + \frac{1}{2} \cdot \frac{1}{2} \cdot \frac{1}{2} \cdot \frac{1}{2} + \frac{1}{2} \cdot \frac{1}{2} \cdot \frac{1}{2} \cdot \frac{1}{2} + \frac{1}{2} \cdot \frac{1}{2} \cdot \frac{1}{2} \cdot \frac{1}{2} + \\
 + \frac{1}{2} \cdot \frac{1}{2} \cdot \frac{1}{2} \cdot \frac{1}{2} + \frac{1}{2} \cdot \frac{1}{2} \cdot \frac{1}{2} \cdot \frac{1}{2}) \Rightarrow K(13) +  < y{ > ^2} = \frac{1}{3}
\end{array}}
 \end{equation}

 \begin{equation}
 \label{eq:47}{\begin{array}{l}
 \bullet K(14) +  < y{ > ^2} = \frac{2}{3}(\frac{1}{2} \cdot \frac{1}{2} \cdot \frac{1}{2} \cdot \frac{1}{2} + \frac{1}{2} \cdot \frac{1}{2} \cdot \frac{1}{2} \cdot \frac{1}{2} + \frac{1}{2} \cdot \frac{1}{2} \cdot \frac{1}{2} \cdot \frac{1}{2} + \frac{1}{2} \cdot \frac{1}{2} \cdot \frac{1}{2} \cdot \frac{1}{2} + \frac{1}{2} \cdot \frac{1}{2} \cdot \frac{1}{2} \cdot \frac{1}{2} + \frac{1}{2} \cdot \frac{1}{2} \cdot \frac{1}{2} \cdot \frac{1}{2} + \\
 + \frac{1}{2} \cdot \frac{1}{2} \cdot \frac{1}{2} + \frac{1}{2} \cdot \frac{1}{2} \cdot \frac{1}{2} \cdot \frac{1}{2} + \frac{1}{2} \cdot \frac{1}{2} \cdot \frac{1}{2} \cdot \frac{1}{2} + \frac{1}{2} \cdot \frac{1}{2} \cdot \frac{1}{2} \cdot \frac{1}{2} + \frac{1}{2} \cdot \frac{1}{2} \cdot \frac{1}{2} \cdot \frac{1}{2}) \Rightarrow K(14) +  < y{ > ^2} = \frac{1}{2}
\end{array}}
 \end{equation}

 \begin{equation}
 \label{eq:48}{\begin{array}{l}
 \bullet K(15) +  < y{ > ^2} = \frac{2}{3}(\frac{1}{2} \cdot \frac{1}{2} \cdot \frac{1}{2} \cdot \frac{1}{2} + \frac{1}{2} \cdot \frac{1}{2} \cdot \frac{1}{2} \cdot \frac{1}{2} + \frac{1}{2} \cdot \frac{1}{2} \cdot \frac{1}{2} \cdot \frac{1}{2} + \frac{1}{2} \cdot \frac{1}{2} \cdot \frac{1}{2} \cdot \frac{1}{2} + \frac{1}{2} \cdot \frac{1}{2} \cdot \frac{1}{2} \cdot \frac{1}{2} + \frac{1}{2} \cdot \frac{1}{2} \cdot \frac{1}{2} \cdot \frac{1}{2} + \\
 + \frac{1}{2} \cdot \frac{1}{2} \cdot \frac{1}{2} \cdot \frac{1}{2} + \frac{1}{2} \cdot \frac{1}{2} \cdot \frac{1}{2} \cdot \frac{1}{2} + \frac{1}{2} \cdot \frac{1}{2} \cdot \frac{1}{2} \cdot \frac{1}{2}) \Rightarrow K(15) +  < y{ > ^2} = \frac{1}{3}
\end{array}}
 \end{equation}

 \begin{equation}
 \label{eq:49}{\begin{array}{l}
 \bullet K(16) +  < y{ > ^2} = \frac{2}{3}\left[ {1 - {{\left( {\frac{1}{2}} \right)}^5}} \right] = \frac{{31}}{{3 \cdot 16}} \Rightarrow \\
 \Rightarrow K(16) = \frac{{29}}{{144}} \approx \frac{1}{{4.9655}}
\end{array}}
\end{equation}

We thus announce our main results as Theorem I :\\
{\bf Theorem I}: The symbol-to-symbol correlation function of the
logistic takes the following values at the Feigenbaum point

\begin{equation}
 \label{eq:50}{{K_{un}}(m) = \left\{ {\begin{array}{*{20}{c}}
{ - \frac{1}{9},}&{m - odd = 1 + 2 \cdot k}&{(k = 0,1,2,...)}\\
{\frac{1}{{18}},}&{m = 2 + 4 \cdot k}&{(k = 0,1,2,...)}\\
{\frac{{10}}{{72}},}&{m = 4 + 8 \cdot k}&{(k = 0,1,2,...)}\\
{\frac{{13}}{{72}},}&{m = 8 + 16 \cdot k}&{(k = 0,1,2,...)}\\
{\frac{{29}}{{144}},}&{m = 16 + 32 \cdot k}&{(k = 0,1,2,...)}
\end{array}} \right.}
\end{equation}

 We can extend this procedure to infinity. The constructive scheme guarantees that this deep algorithmic structure is kept in all
 scales.

The behavior of the symbol-to-symbol correlation function of the
logistic map immediately right to the FP, is an interesting problem
which requires very long trajectories and it is a challenge for future work. \\
As a closing remark, we see that the transition to chaos through
 the period doubling cascade is a phenomenon reflected to  many
 aspects of dynamical systems and to many observables.

 \section{Conclusions }

 The study of correlation functions plays an important role in non-linear science, as it is a measure to characterize the decay of
 correlations in these systems. Motivated by the call of Ruelle and related works [Alonso et al., 1996; May, 1976; Nicolis \& Nicolis, 1988],
we stalled numerical computations to
 study the correlation functions of the logistic map. From all of the correlation functions of the logistic map, correlation function
 at the onset of chaos (Feigenbaum point) plays an important role.\\
In order to focus our attention to computable observables, inspired
by related work by D. Daems and G. Nicolis at the onset of
homoclinic chaos, we rigorously introduced the symbol-to-symbol
correlation function. After checking carefully the arithmetic values
of the correlation function, we obtain a general empirical rule for
its description. \\ Based on the framework of the Metropolis-Stein
and Stein algorithm, we obtain the exact symbolic dynamics at the
accumulation point. This construction allows analytical computations
to obtain the first values of the symbol-to-symbol correlation
function. At this point
 the algorithm connecting the block entropies by lumping and gliding, established
 in [Karamanos \& Nicolis , 1999] leads to a new constructive, step-by-step
 scheme (algorithm) to obtain analytically all values of symbol-to-symbol
 correlation function until infinity.\\
This result gives birth to a
 very strong universality argument for the symbol-to symbol
 correlation function as it holds for any unimodal map
 (it holds for instance for the maps described in [Fraser \& Kapral, 1985]  )
What is the relation of the correlation function of the trajectory,
with the symbol-to-symbol correlation function ${K_{un}}(m)$. A
close relation seems to hold indeed, which is described in a
forthcoming manuscript by the same authors. Roughly, the answer is
that the correlation function of the trajectory contains the same time scales as the symbol-to-symbol correlation function.\\
Thus the blurred image (due to the inherent spatial inhomogeneity of
the corresponding attractor) observed at the trajectory level of
description, is replaced by a set of clear-cut rules at the level of
symbolic dynamics. It is hoped that this study will contribute to a
better understanding of non-linear phenomena.\\

 Acknowledgments{}

The authors express their gratitude to the Library of the National
Research Foundation of Athens, Greece.

\appendix{}
We shall prove that in the space of patterns P, with $P = {\widehat
H^q}(R)$ the following proposition holds :

\text{\bf Proposition A}: The subsequence defined by the odd
(2$\rho$+1)terms of the Feigenbaum sequence is composed by
identically juxtaposed ${2^\rho }$ R-terms.

\textbf{Proof}: The patterns present the particularity that can be
written in the form, P = R$\alpha$R (where $\alpha$ is  substring
composed by
symbols L, R), as can easily be seen by induction. Also,$\widehat {\rm H}(P) = R\alpha R\mu R\alpha R$ .\\
We now define an operator $\widehat G$ which keeps only the odd terms (indices) of the pattern.\\
${1^{st}}$ step: $R \to RLR$

\begin{equation}
 \label{eq:A.1}{\widehat G(RLR) = RR}
\end{equation}

${2^{nd}}$ step :
\begin{equation}
 \label{eq:A.2}{\widehat G(RLRRRLR) = RRRR}
\end{equation}

Suppose that for the ${n^{th}}$ step:

\begin{equation}
 \label{eq:A.3}{\widehat G({\widehat H^n}(R)) = \underbrace {RR......RR}_{{2^n}}}
\end{equation}
Then at the ${(n + 1)^{th}}$ step we have:

\begin{equation}
 \label{eq:A.4}{\widehat G({\widehat H^{n + 1}}(R)) = \widehat G({\widehat H^n}(R){\mu _{{{\widehat H}^n}(R)}}{\widehat H^n}(R)) = \widehat G({\widehat H^n}(R){\mu _{{{\widehat H}^n}(R)}})\widehat G({\widehat H^n}(R)) = \underbrace {RR....RR}_{{2^n}}\underbrace {RR...RR}_{{2^n}}}
\end{equation}
where:
  $\sharp$ of R's at the ${n^{th}}$ step =${2^n}$ = $\sharp$ of odd elements

\begin{equation}
 \label{eq:A.5}{\begin{array}{*{20}{c}}
  {d({{\widehat H}^n}(R)) = (1 + 1) \cdot {2^n} - 1 = {2^{n + 1}} - 1}&{(by}&{induction)}
\end{array}}
\end{equation}

\begin{equation}
 \label{eq:A.6}{\sharp of  R's =\frac{{{2^{n + 1}} - 2}}{2} + 1 = {2^n} - 1 + 1 = {2^n}}
\end{equation}

Secondly we define an operator $\widehat F$, again in the space of
patterns $P$ with  $P = {\widehat H^q}(R)$ which keeps only
the even terms (indices) of the pattern.\\
${1^{st}}$ step: $R \to RLR$
\begin{equation}
 \label{eq:A.7}{\widehat F(RLR) = L = \overline R}
\end{equation}
${2^{nd}}$ step: $RLR \to RLRRRLR$

\begin{equation}
 \label{eq:A.8}{\widehat F({\widehat H^2}(R)) = \widehat F(RLRRRLR) = LRL = \overline {RLR} }
\end{equation}

Let us suppose that for the ${n^{th}}$ step:

\begin{equation}
 \label{eq:A.9}{\widehat F({\widehat H^n}(R)) = \overline {{{\widehat H}^{n - 1}}(R)}}
\end{equation}

Then at the ${(n + 1)^{th}}$ step we have:

\begin{equation}
 \label{eq:A.10}{\begin{gathered}
  \widehat F({\widehat H^{n + 1}}(R)) = \widehat F({\widehat H^n}(R){\mu _{({{\widehat H}^n}(R))}}{\widehat H^n}(R)) = \widehat F({\widehat H^n}(R) \mu  _{({{\widehat H}^n}(R))})\widehat F({\widehat H^n}(R)) =  \hfill \\
{\overline {\widehat H^{n - 1}(R)}}
{ {\overline \mu }  _{({{\widehat H}^n}(R))}}
{\overline {\widehat H^{n - 1}(R)}}=  {\overline{{\widehat H^{n - 1}}(R)}}
{\mu _{({{\widehat H}^{n-1}}(R))} {\overline{{\widehat H^{n - 1}}(R)}} =  {\overline{{\widehat H^{n}}(R)}}} \hfill \\
\end{gathered}}
\end{equation}

Where we have make use of:

\begin{equation}
 \label{eq:A.11}{\overline \mu  ({\widehat H^{n}(R))} = {\mu ({\widehat H^{n - 1}}(R))}}
\end{equation}

the alternation of values of $\mu$'s. In the same manner we arrive at
the following conclusions:

a) The subsequence of the indices of the form ($2\nu  + 4$) is the
displacement of the sequence of even indices ($2\nu  + 2$) for
one place that is:\\

RLLLRLRL$\ldots$     (= $\overline{\mathbb{S}}$)

b) Similarly the subsequence of indices of the form  ($4\nu  + 3$)  is:\\
RRRRRRRR$\ldots$
\\
c)  The subsequence with indices of the form ($2 + 4\nu $) is:\\

LLLLLLLLLL$\ldots$

d) The subsequence of indices of the form ($4 + 8\nu$) is :\\
RRRRRRRRR$\ldots$
\\

\appendix{}

We want to prove that the relation: ${\mathbb{S}_{2 \cdot \nu  +
2}} \circ {\mathbb{S}_{2 \cdot \nu  + 4}} = 0$ is valid for every
$\nu$. For this purpose we will proceed by induction.For this end we
introduce a new quantity:
\begin{equation}
 \label{eq:B.1}{{Q_n} = {\widehat y_{2 \cdot n + 2}} \cdot {\widehat y_{2 \cdot n + 4}}}
\end{equation}

For $n=1$  we have that
\begin{equation}
 \label{eq:B.2}{{Q_1} = {\widehat y_4} \cdot {\widehat y_6} = 1 \cdot 0 = 0}
\end{equation}
If we suppose that
\begin{equation}
 \label{eq:B.3}{{Q_k} = {\widehat y_{2 \cdot k + 2}} \cdot {\widehat y_{2 \cdot k + 4}} = 0}
\end{equation}
is valid, we have to show that is also valid for $k \to k + 1$ e.g

\begin{equation}
 \label{eq:B.4}{{Q_{k + 1}} = {\widehat y_{2 \cdot k + 4}} \cdot {\widehat y_{2 \cdot k + 6}} = 0}
\end{equation}

\begin{equation}
 \label{eq:B.5}{\begin{gathered}
  {\widehat y_{2 \cdot k + 2}} \cdot {Q_{k + 1}} = \underbrace {{{\widehat y}_{2 \cdot k + 2}} \cdot {{\widehat y}_{2 \cdot k + 4}}}_{{Q_k}} \cdot {\widehat y_{2 \cdot k + 6}} = 0 \Rightarrow {Q_k} \cdot {\widehat y_{2 \cdot k + 6}} = 0\mathop  \to \limits^{ \cdot {{\widehat y}_{2 \cdot k + 8}}}  \hfill \\
  \begin{array}{*{20}{c}}
  { \to {Q_k} \cdot {Q_{k + 1}} = 0,}&{\forall k}
\end{array} \hfill \\
\end{gathered}}
\end{equation}


\end{document}